# On thermal boundary layers on a flat plate subjected to a variable heat flux

Jian-Jun SHU and Ioan POP
School of Mechanical & Aerospace Engineering,
Nanyang Technological University, 50 Nanyang Avenue,
Singapore 639798
mjjshu@ntu.edu.sg

*Abstract*—The problem of a steady forced convection thermal boundary-layer past a flat plate with a prescribed surface heat flux is investigated both analytically and numerically. In view of the present formulation, the governing equations reduce to the well-know Blasius similarity equation and to the full boundary-layer energy equation with two parameters: Prandtl number and surface heat flux parameter. The range of existence of solutions is considered. The asymptotic solutions are derived and compared with the numerical solutions of the full boundary-layer equation. A very good agreement between these asymptotic solutions and numerical simulations are found in the range of Prandtl numbers considered.

## I. INTRODUCTION

Historically, the theoretical description of the thermal boundary-layer flow began with the analysis of Prandtl [1], who applied the boundary-layer concept to the heat transfer problems. The work of Prandtl was especially noteworthy as it first introduced the mathematical technique of boundary-layer theory into the subject of heat transfer. Subsequently, Pohlhausen [2] identified similarity solutions for the heat transfer part of the forced convection flow past a flat plate by introducing the dimensionless similarity profile for the boundary-layer energy equation. Pohlhausen's analysis has been very much refined and generalized since then.

It is well-established that convective heat transfer depends on the form of the thermal boundary conditions imposed, with it being usual to take either a prescribed temperature or a prescribed heat flux on the boundary surface. However, in many problems, particularly those involving the cooling of electrical and nuclear components, the wall heat flux is known. In such problems, overheating, burnout, and meltdown are very important issues; therefore, one of the objects of heat transfer theory is the prediction of the wall temperature as wall heat flux varies. The design objective is to control this wall temperature distribution [3]. However, the situation of a prescribed heat flux rate at a surface is often approximated in practical applications and is easier to measure in a laboratory than the case of a surface with prescribed wall temperature.

The main objective of this paper is to complete the existing solutions in the open literature of the forced convection thermal boundary-layer on a flat plate by considering the case when a prescribed wall heat flux is given and is of the form

$$\left(\frac{\partial \theta}{\partial y}\right)_{y=0} = -(1+x^2)^m, \qquad (1)$$

where $x$ and $y$ are the non-dimensional Cartesian coordinates along and normal to the plate, respectively, $\theta$ is the non-dimensional temperature and $m$ is a constant. The analogous problem of free convection boundary-layer flow on a vertical plate immersed in a viscous (Newtonian) fluid with the prescribed wall heat flux given by (1) was discussed by Merkin and Mahmood [4]. In what follows, the solution of the present classical problem with the case given by (1) will be presented and compared with the corresponding results for the prescribed uniform wall heat flux case. It was shown that $m > -\frac{1}{2}$ was a must for a solution of the energy equation to exist. However, it is found that this equation has a physically acceptable asymptotic solution for large $x$ when $m \leq -\frac{1}{2}$. The behaviour of this asymptotic solution for $m \leq -\frac{1}{2}$ is fully discussed. Finally, the analytical solutions for the various values of the parameter $m$ and Prandtl number $P_r$ are compared with the results obtained using numerical techniques. The numerical results confirm the heat transfer features anticipated by the asymptotic solutions.

It is worth mentioning to this end that the precise form that the surface heat flux takes is not important, only that it has the functional forms for small and large $x$ given by (1).

## II. BASIC EQUATIONS

Consider the laminar forced convection boundary-layer flow of a viscous incompressible fluid past a flat plate with a uniform free stream velocity $U_\infty$ and constant temperature $T_\infty$. The boundary-layer equations in terms of the stream function $\psi$ and temperature $\theta$ can be written in non-dimensional form as

$$\frac{\partial \psi}{\partial y}\frac{\partial^2 \psi}{\partial x \partial y} - \frac{\partial \psi}{\partial x}\frac{\partial^2 \psi}{\partial y^2} = \frac{\partial^3 \psi}{\partial y^3}, \qquad (2)$$

$$\frac{\partial \psi}{\partial y}\frac{\partial \theta}{\partial x} - \frac{\partial \psi}{\partial x}\frac{\partial \theta}{\partial y} = \frac{1}{P_r}\frac{\partial^2 \theta}{\partial y^2}, \qquad (3)$$

where $P_r$ is the Prandtl number. Equations (2) and (3) are subject to the boundary conditions

$$y = 0: \quad \psi = \frac{\partial \psi}{\partial y} = 0, \quad \frac{\partial \theta}{\partial y} = -(1+x^2)^m,$$

$$y \to \infty: \quad \frac{\partial \psi}{\partial y} \to 1, \quad \theta \to 0, \qquad (4)$$

where $m$ is a real constant. From (4) it is seen that for small $x$, $\left.\frac{\partial \theta}{\partial y}\right|_{y=0} = -1$, while, for large $x$, $\left.\frac{\partial \theta}{\partial y}\right|_{y=0} = -x^{2m}$, so that though it is possible to write down similarity equations for both small and large $x$, in the latter case these possess a solution only if $m > -\frac{1}{2}$.

A $x$-dependency for $\theta$ is introduced into continuous transformation as follows,

$$\psi = \xi^{\frac{1}{4}} f(\eta), \quad \theta = \xi^{\frac{1}{4}} r_m(\xi) g(\xi,\eta), \quad \xi = x^2, \quad \eta = \frac{y}{\sqrt{x}}. \qquad (5)$$

Without loss of generality it can prescribe

$$r_m(\xi) = (1+\xi)^{m - \left(\frac{1}{2}+m\right)v(-1-2m)}\left[1 + \frac{1}{2}\delta(1+2m)\ln(1+\xi)\right], \qquad (6)$$

where

$$v(z) = \begin{cases} 0, & z < 0, \\ 1, & z \geq 0, \end{cases} \quad \text{and} \quad \delta(z) = \begin{cases} 0, & z \neq 0, \\ 1, & z = 0, \end{cases} \qquad (7)$$

are called the alternative unit-step function and the Kronecker delta function respectively [5]. Substituting (5) into (2) and (3), we get

$$\frac{d^3 f}{d\eta^3} + \frac{1}{2} f \frac{d^2 f}{d\eta^2} = 0, \qquad (8)$$

$$\frac{1}{P_r}\frac{\partial^2 g}{\partial \eta^2} + \frac{1}{2} f \frac{\partial g}{\partial \eta} - \left\{\frac{1}{2} + 2\xi \frac{d}{d\xi}[\ln r_m(\xi)]\right\}\frac{df}{d\eta} g = 2\xi \frac{df}{d\eta}\frac{\partial g}{\partial \xi}, \qquad (9)$$

along with the boundary conditions

$$\eta = 0: \quad f = \frac{df}{d\eta} = 0, \quad \frac{\partial g}{\partial \eta} = -\frac{(1+\xi)^m}{r_m(\xi)},$$

$$\eta \to \infty: \quad \frac{df}{d\eta} \to 1, \quad g \to 0. \qquad (10)$$

The wall temperature distribution is given now by

$$\theta_w(x) = \sqrt{x}\, r_m(x^2) g(x,0). \qquad (11)$$

A standard Keller box method can be adapted to solve (8) and (9) numerically [6-16]. The similarity solutions of (9) are possible for $x = 0$, and are given by

$$\frac{\partial^2 g_0}{\partial \eta^2} + \frac{1}{2} P_r f \frac{\partial g_0}{\partial \eta} - \frac{1}{2} P_r \frac{df}{d\eta} g_0 = 0, \qquad (12)$$

$$\left.\frac{dg_0}{d\eta}\right|_{\eta=0} = -1, \quad g_0|_{\eta \to \infty} \to 0, \qquad (13)$$

which correspond to the thermal boundary-layer on a flat plate with a prescribed uniform heat flux [3]. On the other hand, on letting $x \to \infty$ in (9), we get

$$\frac{\partial^2 g_1}{\partial \eta^2} + \frac{1}{2} P_r f \frac{\partial g_1}{\partial \eta} - \left(\frac{1}{2} + 2m\right) P_r \frac{df}{d\eta} g_1 = 0, \qquad (14)$$

$$\left.\frac{dg_1}{d\eta}\right|_{\eta=0} = -1, \quad g_1|_{\eta \to \infty} \to 0. \qquad (15)$$

These equations correspond to a prescribed wall heat flux $\left.\frac{\partial \theta}{\partial y}\right|_{y=0} = -x^{2m}$ ([17], for $m \geq 0$). By integrating (14) and using (10) for $f$ and (15) for $g_1$, we obtain

$$(1+2m)\int_0^\infty \frac{df}{d\eta} g_1 d\eta = \frac{1}{P_r}. \qquad (16)$$

Hence, we must have

$$m > -\frac{1}{2} \qquad (17)$$

for a solution of (9) to exist for both small and large $x$. However, there are three separate cases to consider for the present problem for large $x$, namely, $m > -\frac{1}{2}$, $m = -\frac{1}{2}$ and $m < -\frac{1}{2}$, respectively.

### III. ASYMPTOTIC SOLUTION FOR $m > -\frac{1}{2}$

The wall temperature distribution for large $x$ is given by

$$\theta_w(x) = g(x,0)\sqrt{x}(1+x^2)^m\left[1 + O\left(\frac{1}{x^2}\right)\right], \qquad (18)$$

where the numerical results of $g_1(0)$ are shown in Table 1 for the different values of the parameter $m$ and Prandtl number $P_r$.

The variation of the wall temperature $\theta_w(x)$ is shown in Fig. 1 for $P_r = 1$. It can be seen that the numerical solution attains its asymptotic condition as given by (18) for the case $m = -\frac{1}{4}$, which corresponds to a uniform wall temperature for large $x$.

TABLE I.   VALUES OF $g_1(0)$ OBTAINED FROM (9)

| $P_r$ | $g_1(0)$ | |
|---|---|---|
| | $m = 0$ | $m = -\frac{1}{4}$ |
| 0.001 | 9.85265 | 9.95016 |
| 0.01 | 8.74748 | 9.52395 |
| 0.1 | 4.93984 | 6.79432 |
| 1 | 2.17879 | 3.01153 |
| 10 | 1.00212 | 1.37336 |
| 100 | 0.46469 | 0.63620 |
| 1000 | 0.21567 | 0.29524 |

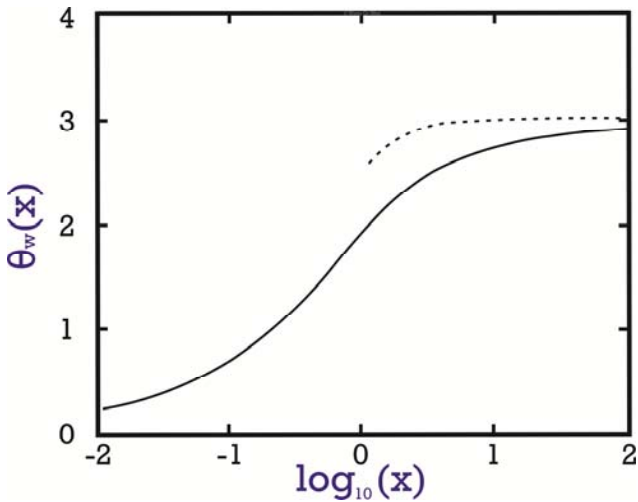

Figure 1.  Graph of $\theta_w(x)$ calculated from the numerical solution of (9) (full line) and from (18) (broken line) for $m = -\frac{1}{4}$ and $P_r = 1$

### IV.   ASYMPTOTIC SOLUTION FOR $m = -\frac{1}{2}$

To find an asymptotic solution of (9), which is valid for large $x$ when $m = -\frac{1}{2}$, we make the transformation

$$\psi = \sqrt{x} f(\eta), \quad \theta = \frac{\ln x}{\sqrt{x}} h(x,\eta), \quad \eta = \frac{y}{\sqrt{x}}, \tag{19}$$

which is the one that gives (14) for the critical case $m = -\frac{1}{2}$.

Using (19), (3) and (4) become

$$\frac{1}{P_r} \frac{\partial^2 h}{\partial \eta^2} + \frac{1}{2} f \frac{\partial h}{\partial \eta} + \left(\frac{1}{2} - \frac{1}{\ln x}\right) \frac{df}{d\eta} h = x \frac{\partial h}{\partial x} \tag{20}$$

subject to the boundary conditions

$$\left.\frac{\partial h}{\partial \eta}\right|_{\eta=0} = -\frac{1}{\ln x} \frac{x}{\sqrt{1+x^2}}, \quad h\big|_{\eta \to \infty} \to 0. \tag{21}$$

We now look for a solution of (20) subject to (21) by expanding $h$ in the form of series

$$h(x,\eta) = h_0(\eta) + O\left(\frac{1}{\ln x}\right), \tag{22}$$

where $h_0$ satisfies the equation

$$\frac{dh_0}{d\eta} + \frac{1}{2} P_r f h_0 = 0, \quad \text{with} \quad h_0\big|_{\eta \to \infty} \to 0. \tag{23}$$

We integrate (3) and apply (4) to get

$$\int_0^\infty \theta \frac{\partial \psi}{\partial y} dy = \frac{1}{P_r} \ln\left(x + \sqrt{x^2+1}\right) = \frac{1}{P_r} \ln x + \frac{1}{P_r} \ln 2 + O\left(\frac{1}{x^2}\right). \tag{24}$$

The asymptotic expression for the wall temperature distribution for large $x$ and $m = -\frac{1}{2}$ can then be expressed as

$$\theta_w(x) = \frac{1}{P_r J} \frac{\ln x}{\sqrt{x}} + O\left(\frac{1}{\sqrt{x}}\right), \tag{25}$$

where the values of $J$ are listed in Table 2 for different Prandtl number $P_r$.

TABLE II.   VALUES OF $J$

| $P_r$ | $J$ |
|---|---|
| 0.001 | 56.03064 |
| 0.01 | 17.66650 |
| 0.1 | 5.45053 |
| 1 | 1.50576 |
| 10 | 0.34919 |
| 100 | 0.07596 |
| 1000 | 0.01638 |

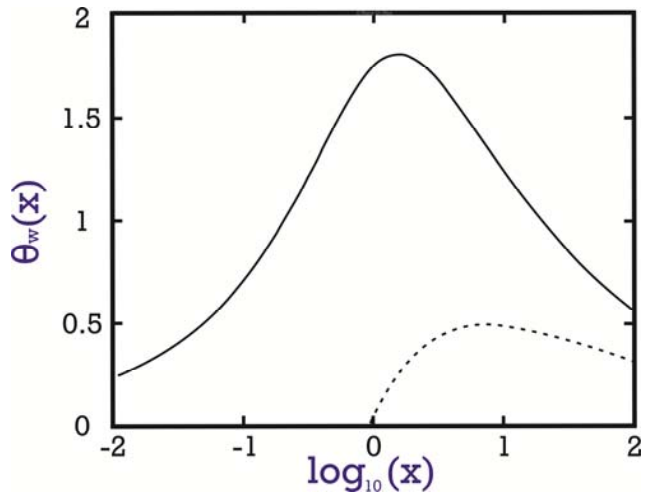

Figure 2.  Graph of $\theta_w(x)$ calculated from the numerical solution of (9) (full line) and from (25) (broken line) for $m = -\frac{1}{2}$ and $P_r = 1$

The variation of the wall temperature $\theta_w(x)$ is shown in Fig. 2 for this case $m = -\frac{1}{2}$ and $P_r = 1$.

## V. Asymptotic solution for $m < -\frac{1}{2}$

To find an asymptotic solution of (9), which is valid for large $x$ when $m < -\frac{1}{2}$, we make the transformation

$$\psi = \sqrt{x} f(\eta), \quad \theta = \frac{1}{\sqrt{x}} G(x,\eta), \quad \eta = \frac{y}{\sqrt{x}}. \quad (26)$$

Using (26), (3) and (4) become

$$\frac{1}{P_r}\frac{\partial^2 G}{\partial \eta^2} + \frac{1}{2} f \frac{\partial G}{\partial \eta} + \frac{1}{2}\frac{df}{d\eta} G = x \frac{df}{d\eta}\frac{\partial G}{\partial x}, \quad (27)$$

subject to the boundary conditions

$$\left.\frac{\partial G}{\partial \eta}\right|_{\eta=0} = -x^{1+2m}\left(1 + \frac{1}{x^2}\right)^m, \quad G\big|_{\eta\to\infty} \to 0. \quad (28)$$

We look for a solution of (27) subject to (28) by expanding $G$ in powers of $x^{1+2m}$ ($1+2m < 0$), of the form

$$G(x,\eta) = G_0(\eta) + O(x^{1+2m}), \quad (29)$$

where $G_0$ satisfies the equation

$$\frac{dG_0}{d\eta} + \frac{1}{2} P_r f G_0 = 0, \quad \text{with} \quad G_0\big|_{\eta\to\infty} \to 0. \quad (30)$$

We integrate (3) and apply (4) to get

$$\int_0^\infty \theta \frac{\partial \psi}{\partial y} dy = \frac{x}{P_r} F\left(-m, \frac{1}{2}; \frac{3}{2}; -x^2\right) = \frac{\sqrt{\pi}\Gamma\left(-m-\frac{1}{2}\right)}{2 P_r \Gamma(-m)} + O(x^{1+2m}) \quad (31)$$

where $\Gamma(z)$ and $F(a,b;c;z)$ are the Gamma function and the Gauss function respectively [5].

The asymptotic expression for the wall temperature distribution for large $x$ and $m < -\frac{1}{2}$ can then be expressed as

$$\theta_w(x) = \frac{\sqrt{\pi}\Gamma\left(-m-\frac{1}{2}\right)}{2 P_r \Gamma(-m) J}\frac{1}{\sqrt{x}} + O\left(x^{2m+\frac{1}{2}}\right). \quad (32)$$

A graph of (32) is shown in Fig. 3 for $P_r = 1$ where we can see that it is in very good agreement with the value obtained from the numerical solution of (9). The case of $m = -1$ was treated by Merkin and Mahmood [4] as an exception due to the existence of eigensolutions. No such eigensolution appears in our case because the momentum equation (Blasius equation) is decoupled from the energy equation.

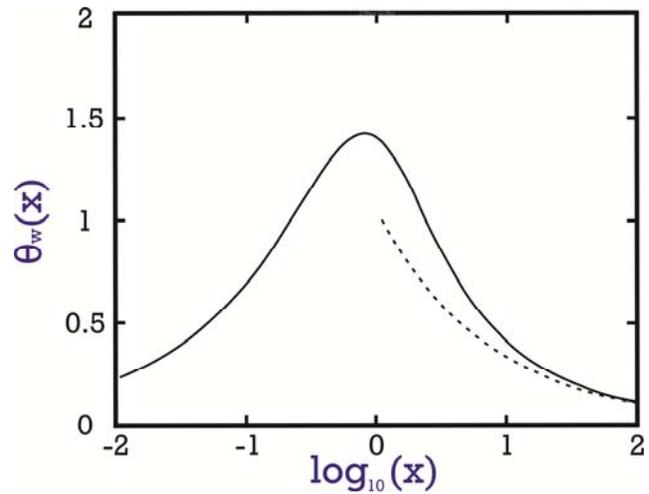

Figure 3. Graph of $\theta_w(x)$ calculated from the numerical solution of (9) (full line) and from (32) (broken line) for $m = -1$ and $P_r = 1$

## Conclusion

The behaviour of the solution of the equations has been considered for the forced convection thermal boundary-layer on a flat plate with a prescribed heating rate proportional to $(1+x^2)^m$. By solving the governing energy equation both analytically and numerically using the Keller box scheme in combination with the continuous transformation method, it has been possible to provide a detailed description of the solutions for large $x$ in the range $m \leq -\frac{1}{2}$ and the different values of the Prandtl number $P_r$. Asymptotic analysis results in an altogether simpler formulation, giving equations whose analytical solution proves to be more easily tractable than the original full boundary-layer equation. Agreement between the asymptotic and numerical solutions for the three regimes $m > -\frac{1}{2}$, $m = -\frac{1}{2}$ and $m < -\frac{1}{2}$ proved to be very good, leading us to believe that the asymptotic approach, although simple in nature, was successful in capturing the essential features of the heat transfer characteristics. The exponent $m$ is, therefore, observed to influence the heat transfer characteristics significantly.

## References

[1] L. Prandtl, "Eine beziehung zwischen wärmeaustausch und strömungswiderstand der flüssigkeiten," Physikalische Zeitschrift, vol. 11, pp. 1072–1078, 1910.

[2] E. Pohlhausen, "Der wärmeaustausch zwischen festen körpern und flüssigkeiten mit kleiner reibung und kleiner wärmeleitung," Zeitschrift für Angewandte Mathematik und Mechanik, vol. 1, pp. 115–121, 1921.

[3] A. Bejan, Convection Heat Transfer, 2nd ed., Wiley, New York, 1995.

[4] J. H. Merkin, and T. Mahmood, "On the free convection boundary layer on a vertical plate with prescribed surface heat flux," Journal of Engineering Mathematics, vol. 24, no. 2, pp. 95–107, 1990.

[5] J. Spanier, and K. B. Oldham, An Atlas of Functions, Hemisphere, Washington, DC, 1987.

[6] J.-J. Shu, and G. Wilks, "An accurate numerical method for systems of differentio-integral equations associated with multiphase flow," Computers & Fluids, vol. 24, no. 6, pp. 625–652, 1995.

[7] J.-J. Shu, and G. Wilks, "Mixed-convection laminar film condensation on a semi-infinite vertical plate," Journal of Fluid Mechanics, vol. 300, pp. 207–229, 1995.

[8] J.-J. Shu, and G. Wilks, "Heat transfer in the flow of a cold, two-dimensional vertical liquid jet against a hot, horizontal plate," International Journal of Heat and Mass Transfer, vol. 39, no. 16, pp. 3367–3379, 1996.

[9] J.-J. Shu, and I. Pop, "Inclined wall plumes in porous media," Fluid Dynamics Research, vol. 21, no. 4, pp. 303–317, 1997.

[10] J.-J. Shu, and I. Pop, "Transient conjugate free convection from a vertical flat plate in a porous medium subjected to a sudden change in surface heat flux," International Journal of Engineering Science, vol. 36, no. 2, pp. 207–214, 1998.

[11] J.-J. Shu, and I. Pop, "Thermal interaction between free convection and forced convection along a vertical conducting wall," Heat and Mass Transfer, vol. 35, no. 1, pp. 33–38, 1999.

[12] J.-J. Shu, "Microscale heat transfer in a free jet against a plane surface," Superlattices and Microstructures, vol. 35, no. 3-6, pp. 645–656, 2004.

[13] J.-J. Shu, and G. Wilks, "Heat transfer in the flow of a cold, axisymmetric vertical liquid jet against a hot, horizontal plate," Journal of Heat Transfer-Transactions of the ASME, vol. 130, no. 1, pp. 012202, 2008.

[14] J.-J. Shu, and G. Wilks, "Heat transfer in the flow of a cold, two-dimensional draining sheet over a hot, horizontal cylinder," European Journal of Mechanics B-Fluids, vol. 28, no. 1, pp. 185–190, 2009.

[15] J.-J. Shu, "Laminar film condensation heat transfer on a vertical, non-isothermal, semi-infinite plate," Arabian Journal for Science and Engineering, vol. 37, no. 6, pp. 1711–1721, 2012.

[16] J.-J. Shu, and G. Wilks, "Heat transfer in the flow of a cold, axisymmetric jet over a hot sphere," Journal of Heat Transfer-Transactions of the ASME, vol. 135, no. 3, pp. 032201, 2013.

[17] E. M. Sparrow, and S. H. Lin, "Boundary layers with prescribed heat flux – application to simultaneous convection and radiation," International Journal of Heat and Mass Transfer, vol. 8, pp. 437–448, 1965.